\documentclass[conference]{IEEEtran}
\IEEEoverridecommandlockouts
\usepackage{cite}
\usepackage{amsfonts}
\usepackage{textcomp}

\usepackage[normalem]{ulem}
\usepackage{amsmath}
\usepackage{amssymb}
\usepackage{multicol}
\usepackage{graphicx}
\usepackage{pifont}
\usepackage[noend]{algorithmic}
\usepackage{algorithm}
\usepackage{enumitem}
\usepackage{multirow}
\usepackage{amsthm}
\usepackage{xcolor}
\usepackage{color}
\usepackage{listings}
\usepackage{subfig}
\usepackage{float}
\usepackage{ragged2e}
\usepackage{xspace}
\usepackage{soul}
\usepackage{lipsum}
\usepackage{makecell}
\usepackage{ulem}
\usepackage{etoolbox}

\newcommand{\proj}{\textsc{SMA\xspace}}
\newcommand{\Fig}[1]{Fig.~\ref{#1}}

\newcommand{\Tbl}[1]{Tbl.~\ref{#1}}
\newcommand{\Sec}[1]{Sec.~\ref{#1}}

\newcommand{\benchmark}[1]{{\texttt{#1}}}

\renewcommand{\paragraph}[1]{\vspace*{.0cm}\noindent\textbf{#1}\hspace*{.1cm}}

\def\BibTeX{{\rm B\kern-.05em{\sc i\kern-.025em b}\kern-.08em
    T\kern-.1667em\lower.7ex\hbox{E}\kern-.125emX}}

\begin{document}

\title{\huge Balancing Efficiency and Flexibility for DNN Acceleration via Temporal GPU-Systolic Array Integration
}


\author{
  \IEEEauthorblockN{
  \small Cong Guo$^1$, Yangjie Zhou$^1$, Jingwen Leng$^{1,*}$, Yuhao Zhu$^2$, Zidong Du$^3$, Quan Chen$^1$, Chao Li$^1$, Bin Yao$^1$, Minyi Guo$^{1,*}\thanks{* Jingwen Leng and Minyi Guo are co-corresponding authors of this paper.}$\\
$^1$\textit{Shanghai Jiao Tong University}, 
$^2$\textit{University of Rochester},
$^3$\textit{Institute of Computing Technology, Chinese Academy of Sciences}
}

}

\makeatletter
\patchcmd{\@maketitle}
  {\addvspace{0.5\baselineskip}\egroup}
  {\addvspace{-1.5\baselineskip}\egroup}
  {}
  {}
\makeatother
  
\maketitle

\begin{abstract} 
The research interest in specialized hardware accelerators for deep neural networks (DNN) spikes recently owing to their superior performance and efficiency. However, today's DNN accelerators primarily focus on accelerating specific ``kernels'' such as convolution and matrix multiplication, which are vital but only part of an end-to-end DNN-enabled application. Meaningful speedups over the entire application often require supporting computations that are, while massively parallel, ill-suited to DNN accelerators. Integrating a general-purpose processor such as a CPU or a GPU incurs significant data movement overhead and leads to resource under-utilization on the DNN accelerators.

We propose Simultaneous Multi-mode Architecture (SMA), a novel architecture design and execution model that offers general-purpose programmability on DNN accelerators in order to accelerate end-to-end applications. The key to SMA is the temporal integration of the systolic execution model with the GPU-like SIMD execution model. The SMA exploits the common components shared between the systolic-array accelerator and the GPU, and provides lightweight reconfiguration capability to switch between the two modes in-situ. The SMA achieves up to 63\%  performance improvement while consuming 23\% less energy than the baseline Volta architecture with TensorCore.
\end{abstract}

\section{Introduction}\label{sec:introduction}
Deep learning has revolutionized the key domains such as natural-language processing and computer vision~\cite{krizhevsky2012imagenet}, for which the GPU is the enabler.
Although still widely used for training and inference, GPU's performance and energy efficiency start to plateau.
As such, computer architects start to build specialized hardware accelerators for deep neural network models, which have high computation intensity and regular dataflow patterns.
 Prime examples include Eyeriss~\cite{7551407}, Diannao Family~\cite{chen2016diannao}, TPU~\cite{jouppi2017datacenter}, and TensorCore (TC)~\cite{volta2017whitepaper}.


However, existing accelerators are insufficient to provide meaningful speedup for DNN-enabled applications. This is because they focus only on specific ``kernels'' such as convolution, but ignore the end-to-end application characteristics.
As a result, accelerating those fixed kernels leads to insignificant speedup, sometimes even slowdown, at the application-level. 
In specific, emerging ``hybrid'' neural network models have started to combine the convolution kernels with irregular kernels that are, although massively parallel, ill-suited for specialized accelerators. Besides, applications like autonomous driving~\cite{lin2018architectural} adopt both the neural network models and algorithms from other domains with different characteristics.

To support the end-to-end applications, specialized accelerators must be coupled with flexibilities for unsupported computations. 
The mainstream solutions fall into three categories, neither being efficient. 
The first approach integrates the accelerator with a general-purpose host processor, which incurs significant data movement overhead and  accelerator resource under-utilization during the unsupported computations. 
Second, systems like TPU could convert the unsupported operations to compatible operations for the native execution, but with the severe efficiency loss.
Finally, the GPU-based systems provide general-purpose programmability for different types of operations, but are less efficient than the specialized accelerators when executing those regular kernels.


We propose \underline{s}imultaneous \underline{m}ulti-mode \underline{a}rchitecture (SMA), which provides flexibility in accelerating the irregular operations while maintaining efficiency in regular convolution operations. It leads to significant end-to-end application speedup compared to state-of-the-art systems. Our design philosophy starts with the SIMD execution in the GPUs, and judiciously apply lightweight architectural augmentations to support the systolic execution model, which is proven efficient for regular operations like convolutions and matrix multiplication~\cite{kung1982systolic}.

Our integration of a specialized accelerator on a general purpose substrate shares the similarity in the recent  tensor core (TC) in the NVIDIA Volta GPU~\cite{raihan2019modeling}.
However, our architecture is different from the tensor core in the two following key aspects. 
First, we \textit{temporally} integrate the systolic array and SIMD architecture while TC does so \textit{spatially} which leads to area wastage. Second, we employ a SIMD-friendly systolic array dataflow, which achieves high data reuse while maximizing the GPU memory access efficiency. In contrast, the dataflow in tensor core suffers from the low data reuse.

The critical challenge in our architecture is to achieve a high execution efficiency in the systolic mode and a low runtime reconfigurability overhead between the systolic mode and SIMD mode. We leverage the GPU's massive parallelism and a fine-grained synchronization scheme, which is able to control the execution of systolic array with little overhead.
The systolic mode eliminates most of the expensive accesses to register file and shared memory, and significantly improves the matrix multiplication efficiency than the SIMD-only mode.
On the other hand, the SIMD mode enables efficient acceleration of hybrid DNN workloads as it preserves the programmability.

The contribution of our work is as follows:

\begin{itemize}[leftmargin=*]

\item We quantify the execution inefficiencies of running DNNs with irregular operations on a variety of existing hardware platforms and identify their execution bottlenecks.


\item  We consider various systolic array dataflows and identify a SIMD-friendly one that  balances memory access efficiency and data reuse to facilitate its integration on GPUs.

\item  We propose \proj{}, an architecture that temporally integrates SIMD and systolic execution model with little reconfiguration overhead by exploiting the architectural similarity.

\end{itemize}


\section{Analysis of Existing DNN Accelerators}
\label{sec:motivation}

We analyze the performance of executing the emerging hybrid DNN models on existing hardware accelerators, i.e., TPU~\cite{jouppi2017datacenter} and TensorCore (TC)~\cite{volta2017whitepaper}.
They are the two commercially available DNN accelerators with highly optimized software stacks.
However, our experimental results identify that the GEMM-incompatible computation in the hybrid models is too computational-demanding to run on the hardware accelerators even coupled with a general-purpose CPU.

   \begin{figure}[t]
    \vspace{-0.5cm}
      \includegraphics[width=1\columnwidth]{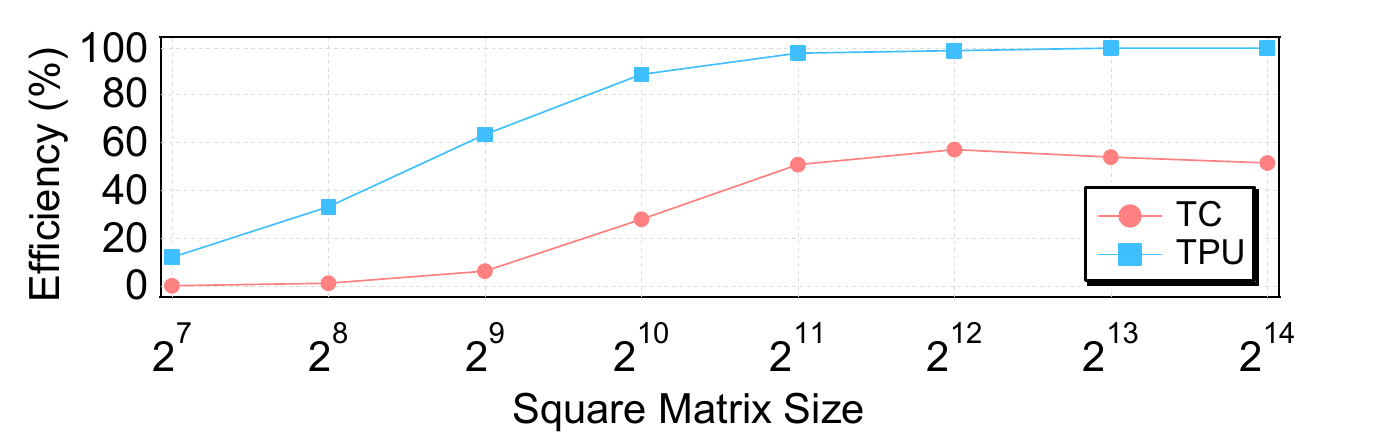}
    \vspace{-0.6cm}
      \caption{Tensor Core and TPU Efficiency.}
      \label{fig:tc_eff}
    \vspace{-0.5cm}
    \end{figure}

\subsection{DNN Hardware Accelerators}

Both TPU and TC accelerate the GEMM operation, which is the dominant operation in the commonly used models, such as convolution neural network~\cite{krizhevsky2012imagenet} (CNN), multi-layer perceptron (MLP), and RNN/LSTM~\cite{mikolov2010recurrent}.

The TPU uses a weight stationary systolic array~\cite{kung1982systolic}, whose size is $256 \times 256$ in TPU-v1. 
The systolic array has a significant degree of data reuse, which leads to high-performance and energy-efficient execution of GEMM.
In contrast, TC is still SIMD architecture and has a limited degree of data reuse.
According to the reverse-engineered work~\cite{raihan2019modeling}, it executes the GEMM operation in the dot-product fashion and supports a $4 \times 4 \times 4$ GEMM operation, much smaller than the TPU.

We compare the GEMM performance between TPU and TC in \Fig{fig:tc_eff}. 
We use a cloud TPU-v2, which has a total of eight cores.
We only use one core that has a $128 \times 128$ systolic array with peak $22.5$ TFLOPS.
The GPU is a Tesla V100, which has 15.7 FP32 and 125 TC peak TFLOPS~\cite{volta2017whitepaper}.
Owing to their different peak FLOPS,  we use the FLOPS efficiency (achieved FLOPS divided by the peak FLOPS)  as the metric for a fair comparison.
With a large enough matrix, the TPU achieves almost 100\% FLOPS efficiency, while the TC achieves less than 60\% efficiency.
As previously explained, the TC calculates the GEMM with multiple parallel dot-product operations while the TPU does it in the systolic fashion.
As a result, a tensor core only supports a small GEMM operation with limited data reuse and high register bandwidth consumption, which leads to its low FLOPS efficiency.


\subsection{Hybrid Models}
\label{sec:hybrid}

We now compare the performance of the two accelerators on hybrid models.
Those models, which are the results of fast-evolving DL algorithms, can have operations that cannot be executed through the GEMM, and therefore present significant challenges for the existing fixed-function accelerators.  

\begin{figure}
    \vspace{-0.8cm}
    \includegraphics[width=\columnwidth]{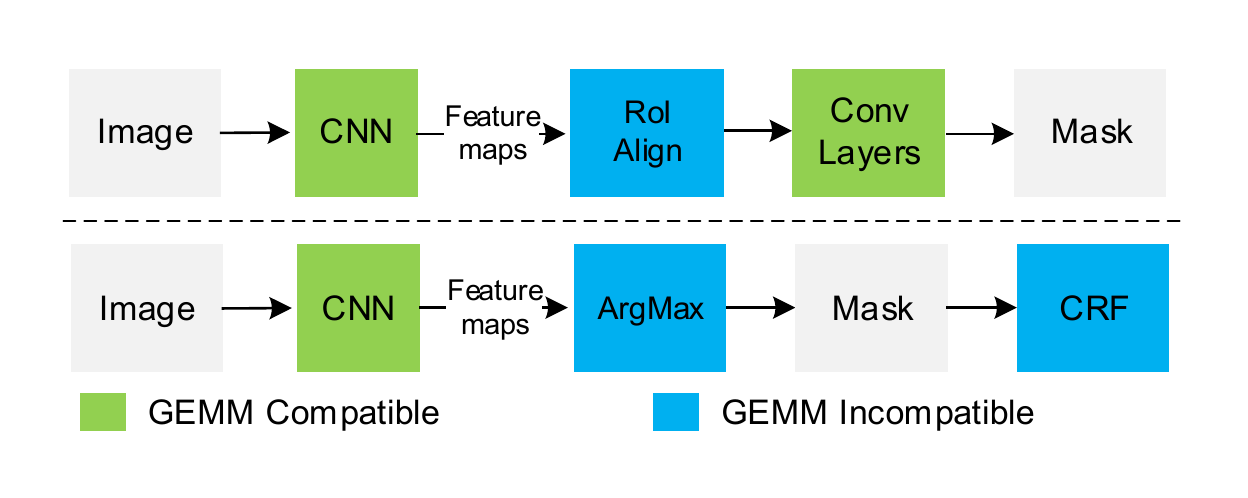}
    \vspace{-0.8cm}
    \caption{Mask R-CNN (top) and DeepLab (bottom) details.}
    \vspace{-0.4cm}
    \label{fig:maskrcnn}
   \end{figure}

\Fig{fig:maskrcnn} shows two such hybrid DNN models, i.e., Mask R-CNN~\cite{he2017mask} and DeepLab~\cite{chen2018deeplab}.
Both models target the semantic segmentation task, which aims to classify all the pixels in the input image and is more complicated than the image classification. 
As such, the state-of-the-art models rely on CNN models for feature extraction, but also introduce additional operations to improve the segmentation accuracy.
Both models in \Fig{fig:maskrcnn} have GEMM-compatible \benchmark{CONV} and \benchmark{FC} layers. But Mask R-CNN has \benchmark{RoIAlign}, a bi-linear interpolation that requires many reshape operations, and \benchmark{RegionProposal}, a control-flow intensive non-max suppression (NMS) algorithm.
Those operations are challenging to support with only GEMM operation.
Similarly, DeepLab has the \benchmark{ArgMax} and \benchmark{CRF} (conditional random field~\cite{lafferty2001conditional}) that are GEMM-incompatible.

\begin{figure}[t]  
  \minipage{0.16\textwidth}
    \includegraphics[height=0.5\linewidth]{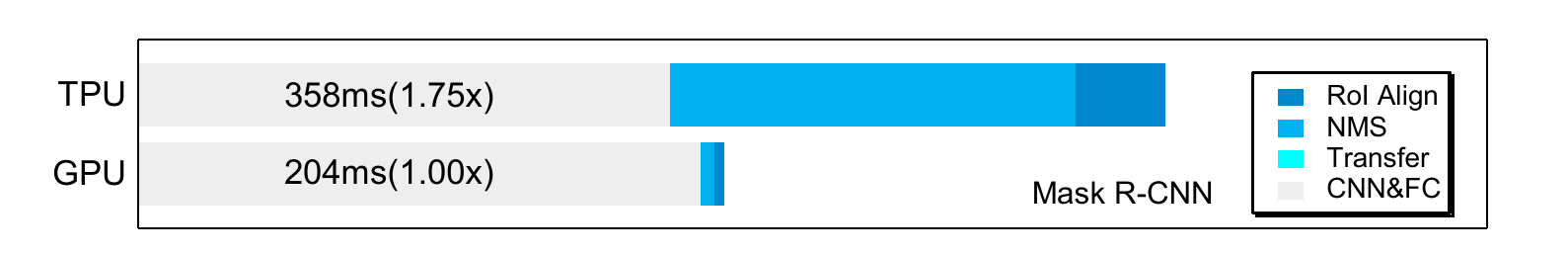}
  \endminipage 
  \vfill  
  \hspace{0mm}
  \minipage{0.16\textwidth}
    \includegraphics[height=0.5\linewidth]{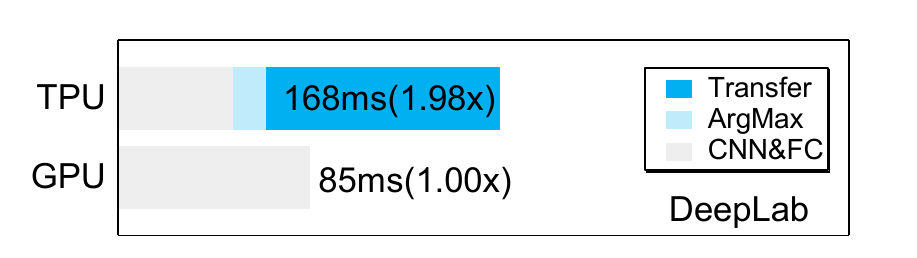}
    \endminipage
    \hfill
    \minipage{0.16\textwidth}
    \includegraphics[height=0.5\linewidth]{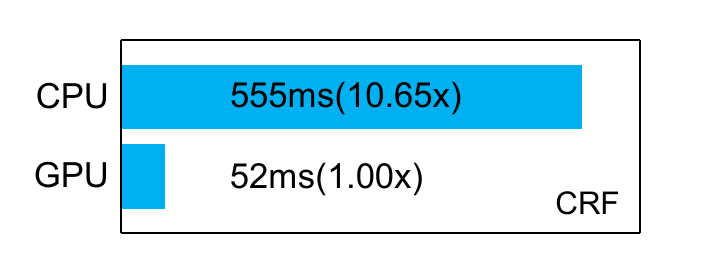}
  \endminipage
  \hspace{10.5mm}
  \caption{TPU vs GPU for Mask R-CNN and DeepLab.}  \label{fig:2results}
  \vspace{-0.6cm}
  \end{figure}
  
\Fig{fig:2results} shows the performance comparison and breakdown on TPU and GPU.
The TPU executes Mask R-CNN about 75\% slower than the GPU. A closer observation reveals that the GPU is slower than the TPU on the GEMM-compatible kernels (i.e., \benchmark{CONV} and \benchmark{FC}), but far out-performs the TPU on \benchmark{RoIAlign} and \benchmark{RegionProposal}. 
We examine the TPU-version source code for performance debugging and find that it converts the control-flow intensive \benchmark{NMS} operation in \benchmark{RegionProposal} to multiple dataflow-based GEMM operations, and converts \benchmark{RoIAlign} operation to multiple average pooling operations for which TPU has hardware support.
As such, the TPU can execute Mask R-CNN without relying on the CPU, i.e., with no data transferring overhead.
However, the improper mapping causes severe performance degradation.

The DeepLab runs much slower on the TPU than the GPU owing to its infeasibility to support the important \benchmark{CRF} operation.
As such, the TPU transfers the data to the CPU for executing \benchmark{CRF}, and we separate the \benchmark{CRF} time from the overall execution time.
The TPU has higher performance ($> 1.6\times$) than the GPU for the GEMM-compatible kernels, but the data transferring overhead is $1.2\times$ of its GEMM operation, leading to an overall $2 \times$ slowdown compared to the GPU.
Also, the performance of \benchmark{CRF} is $10 \times$ worse on the CPU (with one core). 

The results show that over-specialization can severely degrade the performance for incompatible operations so that it is crucial to provide general-purpose programmability for emerging hybrid DNN models.
However, the approach of relying on general-purpose cores can cause significant data movement overhead and also fails to exploit the computation resources inside the accelerator.

\section{Simultaneous Multi-mode Architecture}
\label{sec:architecture}

To balance the efficiency and flexibility, we propose the simultaneous multi-mode architecture. SMA integrates a GPU-like SIMD execution mode and a systolic execution mode, and temporally switches between the two modes.
The SIMD mode efficiently executes GEMM-incompatible operations, and the specialized mode accelerates GEMM operations.
We describe the design principles behind \proj{} and highlight its key novelty over TC, another architecture instance that can switch between generic SIMD execution and DNN acceleration.

\subsection{Temporal Integration}

The first design principle in SMA is the temporal integration between the general-purpose mode and specialized mode. 
In contrast, the TC adopts the spatial integration methodology, which leads to the overhead in both the area and performance.

As explained previously, each TC has multiple dot-product units for executing matrix multiplication, and each dot-product unit has 4 MAC units~\cite{raihan2019modeling}.
The SIMD units in GPU have the same computation ability but are not used when TC is active, which is essentially area wastage.
Meanwhile, the TC also requires an adder tree for result reduction, which incurs additional area overhead.
Spatial integration of the two architectures leads to area inefficiency and resource wastage because the computation of DNN models is usually layer-by-layer where only one architecture is used when performing the computation for one layer.
In contrast, \proj{} is built on top of the existing SIMD execution units and aims for the maximal sharing of hardware structures (i.e., improved area efficiency). 


The spatial integration of TC also leads to its highly decoupled execution model~\cite{appleyard_yokim_2019}: the SIMD units load data to register file and the TC relies on an explicit synchronization to receive the data.
In addition, the TC only supports a fixed shape (i.e., $4 \times 4 \times 4$) of matrix multiplication and does not expose the opportunity of SIMD-accelerator collaboration to more aggressively hiding the data loading latency.
This decoupled execution model has inherent performance inefficiency. 
In contrast, the temporal integration in \proj{} enables such collaboration by imposing zero switching overhead between SIMD and accelerator mode.


\subsection{Choice of Dataflow}
\label{sec:architecture:em}


SMA starts with a SIMD substrate and adds another systolic mode for DNN acceleration because systolic array exploits data reuse and outperforms the dot-product-based TC (\Fig{fig:tc_eff}).
However, the SIMD and systolic array favor distinct memory access patterns, which in turn expose a fundamental trade-off between memory access efficiency and data-reuse that we must reconcile when integrating the two modes.



\begin{figure}[t]
\vspace*{-0.4cm}
  \includegraphics[width=\columnwidth]{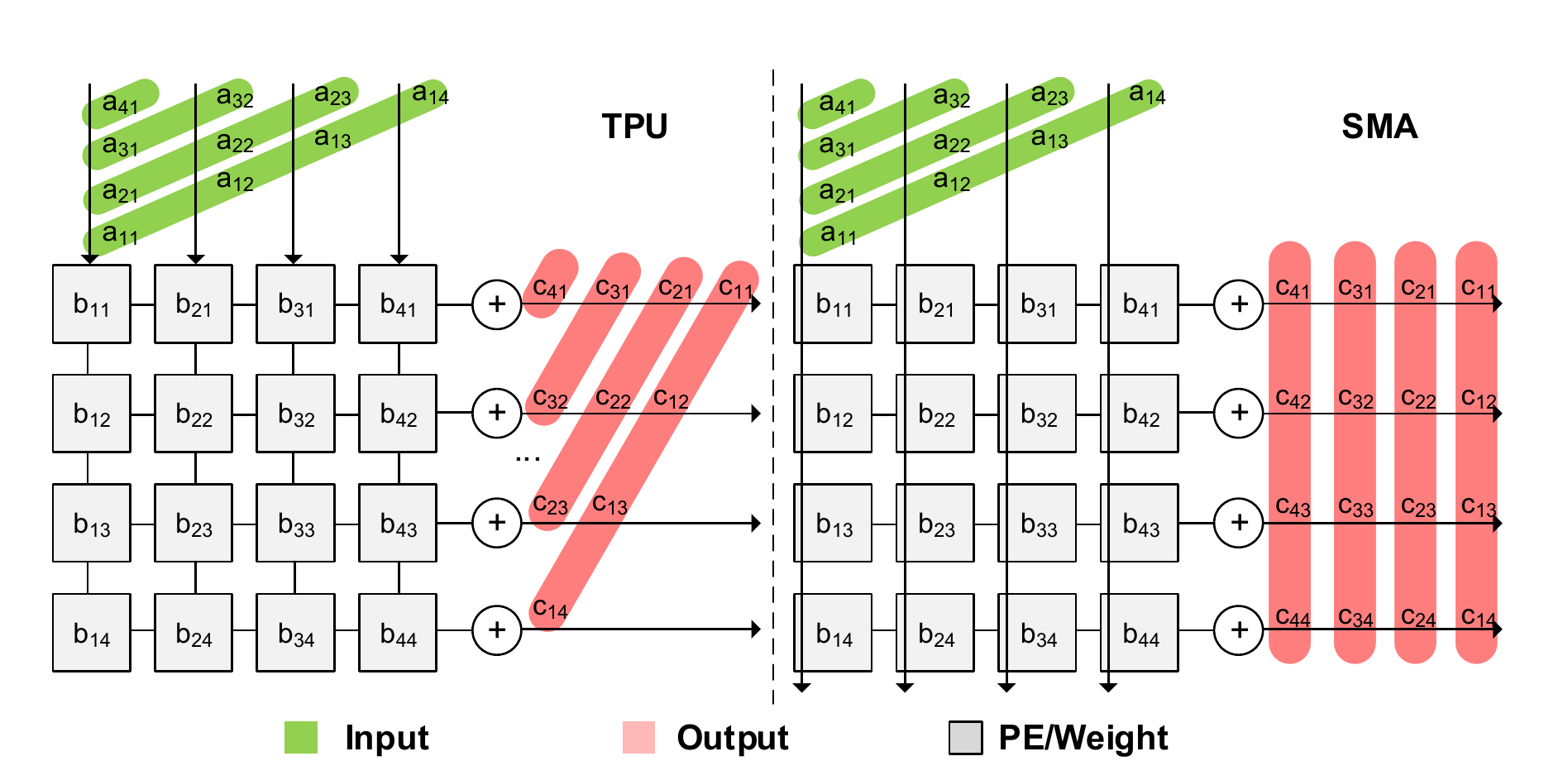}
  \caption{Comparison of TPU and \proj{} dataflow.} 
\vspace{-0.6cm}
  \label{fig:dataflows}
\end{figure}

The SIMD architecture of GPU favors a \emph{coalesced} memory access, which is supported by the cache system~\cite{Leng:2013:GEE:2485922.2485964}. 
 However, systolic array incurs memory accesses that could not be coalesced.
\Fig{fig:dataflows} (left) shows the weight-stationary dataflow in the TPU.
Loading matrix $A$ and writing output matrix $C$ requires accessing different rows of the respective matrix as the same cycle. Thus, directly implementing the systolic execution model on the SIMD hardware would lead to unsupported memory behaviors. While shared memory (scratchpad) in GPUs supports uncoalesced memory accesses via banking, it has a limited number of banks and thus does not scale well to large (or multiple) systolic arrays.

The TC chooses to address the problem by favoring coalesced memory accesses to maximally utilize its GPU's memory subsystems. More specifically, TC uses a set of dot-product units to implement GEMM~\cite{raihan2019modeling}. In that way, all the memory accesses (matrix $A$, $B$, and $C$) are coalesced in TC.
However, this approach leads to low data reuse and therefore poor GEMM performance as evident in \Fig{fig:tc_eff}.

To strike a balance of memory access efficiency and data reuse, we analyze different systolic array dataflow proposed in prior work and identify a SIMD-friendly dataflow called semi-broadcasted weight-stationary~\cite{kung1982systolic}.
The dataflow, shown in \Fig{fig:dataflows} (right), is similar to the TPU's data.
However, instead of passing a matrix $A$ element from top to bottom in the original design, each $A$ element now is broadcasted to all the PEs in the same column. Every cycle, all the PEs in the same column get the same element from matrix $A$, perform a MAC operation, and send the data to the corresponding PEs on the right. 

This dataflow is more SIMD-friendly and enables the seamless integration on a SIMD substrate.
Specifically, each element in the matrix $A$ and $C$ is reused $N$ times in the $N \times N$ array, which is the same as the weight-stationary systolic execution model and is better than the TC. Meanwhile, accesses to matrix $B$ and $C$ are coalesced, and only accesses to matrix $A$ are uncoalesced.

\section{SMA Implementation} 
\label{sec:implementation}

This section describes the architectural modification to the baseline GPU to implement the \proj.
The architectural design challenge is that SMA \textit{temporally} integrates the SIMD mode and the systolic mode, and temporally reconfigures itself at runtime. Thus, we must minimize the reconfiguration overhead with minimal hardware augmentation.
We then explain the instruction control and SIMD-systolic interaction.
The temporal integration is feasible owing to our SIMD-friendly systolic dataflow that has high enough architectural similarity compared to the baseline GPU.
As such, our design can reuse as many existing resources (i.e., computation, memory, and control) and architectural features as possible.


\begin{figure}[t]
  \begin{center}
    \vspace{-6mm}
  \centering
  \includegraphics[width=1\linewidth]{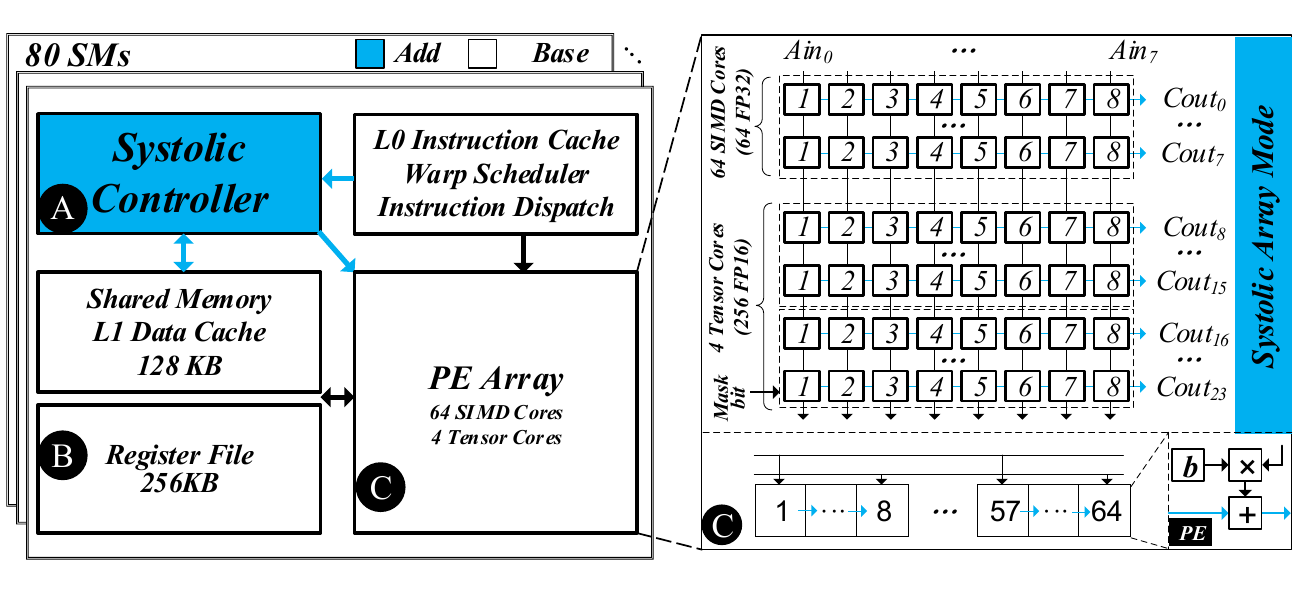}
  \vspace{-6mm}
  \caption{The details of SMA: A. systolic controller; B. shared memory and register file; C. PE array.}
  \vspace*{-0.4cm}
  \label{fig:base_arch}
  \end{center}
  \vspace{-6mm}
\end{figure}

\subsection{SMA Unit Design}

The heart of our design is a set of \proj{} units, with each being a systolic array in the specialized mode and reconfigured to conventional SIMD lanes in the general-purpose mode.

We use the latest Volta architecture~\cite{volta2017whitepaper} in \Tbl{tbl:sma} as our baseline GPU architecture, which has 80 cores (or streaming multiprocessor, SM).
Each SM has 64 CUDA cores (i.e., 64 FP32 unit) and 4 TCs (i.e., 256 FP16 units \textit{in total}).
It also has up to 96~KB shared memory that has 32 banks and each bank provides a 32-bit value.
The register file can provide vector-like access and its size is 256~KB in an SM.

 SMA reuses the same computation resources in each SM (64 CUDA cores and 4 TCs, equivalent to 128 FP32 units in total), and provides three SMA units per SM. Each SMA unit is a $8 \times 8$ FP32 systolic array.
\Fig{fig:base_arch} shows the microarchitecture of one $8\times 8$ \proj{} unit. The SMA unit is implemented on top of the baseline SM architecture in the GPU with two key architectural augmentations to support the semi-broadcasted weight-stationary data-flow (\Sec{sec:architecture:em}). 

First, we repurpose the existing operand collector as a local buffer for storing the stationary weights of each PE. Second, we add additional wires to support broadcasting elements in matrix $A$ and communicating partial sums within the array. Overall, the layout of the $8 \times 8$ systolic array could be done with minimal routing changes to the physical layout of existing SIMD units as the bottom of \Fig{fig:base_arch}(C) shows.


Certain NVIDIA GPUs support the precision conversion between FP32 and FP16~\cite{volta2017whitepaper}.
For example, two FP16 MAC units can be grouped to one FP32 MAC unit.
If the baseline GPU supports this precision conversion, our \proj{} can also exploit it, leading to an $8 \times 16$ FP16 systolic array instead of the current $8 \times 8$ FP32 systolic array. 
Similarly, our \proj{} unit can also be built from other data types such as INT8.

{
\renewcommand{\arraystretch}{1.2}
\begin{table}[b]

  \vspace{-0.4cm}
\centering  
\resizebox{\linewidth}{!}{
\begin{tabular}{l | l l} 
\Xhline{1.5pt} 
& \textbf{GPGPU} & \textbf{\proj{}}\\\hline
\textbf{Baseline} & Volta  & Volta  \\
\hline
\textbf{SMs} & 80 & 80 \\
\hline
\textbf{CUDA Core/SM} & 64 FP32 units & \multirow{2}{*}{3 $8 \times 8$ \proj{} unit}  \\
\cline{1-2}
\textbf{Tensor Core/SM} & 4 (256 FP16 units)  & \\
\hline
\multirow{2}{*}{\textbf{Shared Memory/SM}} & 32 banks & 32 banks (8 for all \proj{} units) \\
&Configurable up to 96KB & Configurable up to 96 KB \\
\hline
{\textbf{Register File/SM}}  & 256 KB & 256 KB  \\
\hline
\Xhline{1.5pt}
\end{tabular}
}
\caption{Baseline GPU and \proj{} Configurations.}
\vspace{-0.2cm}
 \label{tbl:sma} 
\end{table}
}

\subsection{Instruction Control}
\label{subsec:control}

We present our asynchronous instruction based control mechanism that can be seamlessly integrated into the existing SIMD pipeline and enable the simultaneous presence of two distinctive modes.
Under the hood, \proj{} uses GPU's rich memory resources, abundant parallelism, and fine-grained synchronization to maximize the performance.





We propose a new instruction LSMA (Load, Store and Multiply-accumulate) for the systolic mode.
The instruction executes the operation in Eq.~\ref{lsma} and requires four register operands: the addresses of the first element in matrix $A$ and $C$, one element value in matrix $B$, and the height of matrix $A$. The instruction executes asynchronously with respect to other SIMD instructions, minimizing the interference to the existing pipeline control logic.
The threads need to issue an explicit synchronization to access the systolic computation results.
\vspace{-1mm}
\begin{equation}\label{lsma}
    \text{LSMA}\ B \Rightarrow
    C[out]\leftarrow A[in]\times B+C[in]
  \end{equation}
  \vspace{-6mm}

Once LSMA instruction is issued, a dedicated systolic controller in \Fig{fig:base_arch} is responsible for controlling the array with two main roles.
First, it has an active mask for controlling the idle or active status of individual PE.
Second, it has multiple address generation units for feeding the data to the array that has two different kinds of memory accesses (\Sec{sec:architecture:em}).
For an SMA unit, we assign 8 shared memory banks for loading matrix $A$ with uncoalesced accesses and one register file (RF) bank for storing matrix $C$ with coalesced accesses.
In the baseline GPU, a RF bank provides 32 values (32-bit) for a warp, which is enough for the SMA unit that reads 8 values in a cycle.
The three SMA units can be combined as an $8 \times 24$ array to coordinate their memory accesses.
\Fig{fig:base_arch}(C) shows the register/shared memory access for the three SMA units.

TC has inherent inefficiency owing to the strictly synchronous semantics and fixed matrix shape (\Sec{sec:architecture:em}).
In contrast, our instruction design overcomes this inefficiency by adopting the asynchronous semantics and a flexible shape ($K \times 8 \times 8$), which enables the more fine-grained SIMD-systolic collaboration as we describe in the next subsection.

\subsection{Algorithm Mapping}
\label{sec:gemm_impl}

\begin{figure}[t]
  \begin{center}
    \vspace{-7mm}
  \includegraphics[width=0.9\columnwidth]{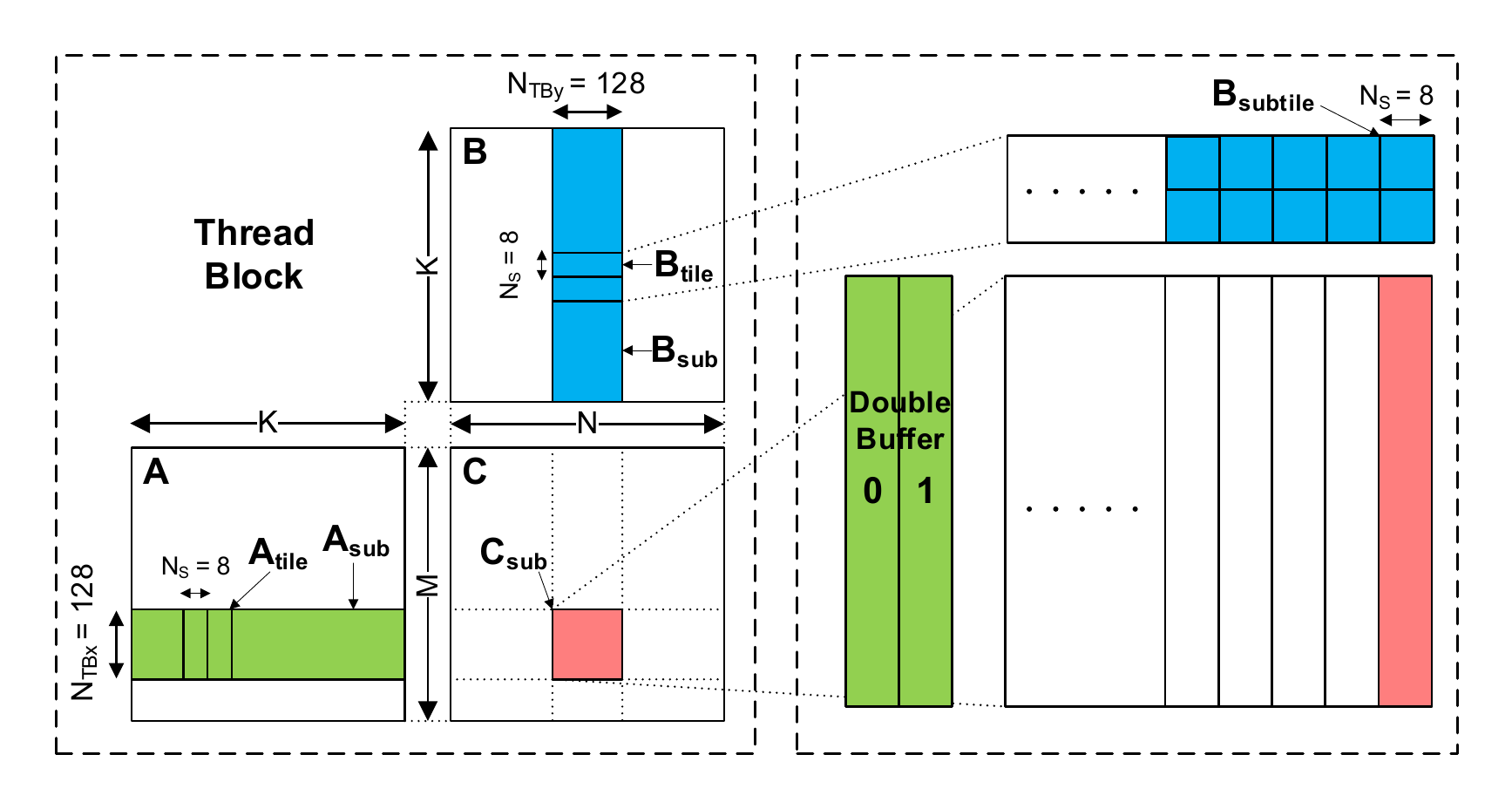}
  \vspace{-2mm}
  \caption{GEMM partition and tiling on SMA.}
  \label{fig:tilting}
  \vspace{-3mm}
  \end{center}
  \vspace{-4mm}
\end{figure}

We describe the algorithm mapping and optimization in \proj{}, most of which run at the software level and leverage the SIMD-systolic collaboration.
We also present the new \proj{}-specific warp synchronization primitive and scheduler.

We implement the GEMM of $C = \alpha A \times B + \beta C$ and adopt common parallelization techniques such as partition, tiling, and double buffering as shown in \Fig{fig:tilting}.
We divide the computation by the output matrix $C$ on the two-dimensional grid of thread blocks (TBs).
This partition avoids the inter-TB communication and each TB is responsible for calculating a sub-matrix of $C_{sub}$, which is stored in the register file for faster access.
Owing to the constraints of register file capacity, we choose the sub-matrix size of $128\times 128$.

For increased data-locality, $A_{sub}$ and $B_{sub}$ are divided into tiles of $A_{tile}$ and $B_{tile}$ with the size of $128\times 8$.
Different tiles work in the double-buffer fashion: in each iteration, each TB uses $A_{tile}$ and $B_{tile}$ to update the sub-matrix $C_{sub}$.
Since each core has an $8 \times 8$ weight-stationary systolic array, we divide the $128 \times 8$ tile $B_{tile}$ into 16 sub-tiles to run on the systolic array.  
As such, each systolic array operation computes $A_{tile}$ ($128 \times 8$) and $B_{subtile}$ ($8 \times 8$).
To maximize the concurrency, we use 64 warps (i.e. 2048 threads) per TB, which are divided into two sets for double buffer. 
The two sets work alternatively between loading tiles with the SIMD mode and computing the tile with the systolic mode via the LSMA instruction. 
The warps in the two sets are synchronized through CUDA's recently added fine-grained sync primitive cooperative groups~\cite{nvidia2019toolkit}.




 
The challenge of running the double-buffered algorithm on the GPU lies in the architecture's throughput-oriented design, which leads to its greedy-then-oldest (GTO) warp scheduler.
The scheduler tries to issue the same set of warps over and over to maximize the throughput, which may cause starvation in the double-buffered warps.
To overcome such a challenge, we add a \proj{}-specific scheduler that works in the round-robin fashion.
The new scheduler works only in the systolic mode and does not affect the original scheduler.


\section{Evaluation}\label{sec:result}

In this section, we perform comprehensive performance and energy efficiency evaluation of \proj{} in different scenarios.
For regular DNN models, we compare \proj{} with its baseline SIMD architecture and demonstrate its efficiency and flexibility for supporting both the regular and hybrid models.
In the end, we evaluate the \proj{}'s dynamic resource scheduling capability in the context of autonomous driving application that contains both DNN and traditional algorithms.

{
\renewcommand{\arraystretch}{1.2}
\begin{table}[b]
\vspace*{-2mm}
\centering  
\resizebox{\linewidth}{!}{
\begin{tabular}{l | l | l | l| l| l} 
\Xhline{1.5pt} 
\textbf{Network} & \textbf{AlexNet} & \textbf{VGG-A} &\textbf{GoogLeNet} & \textbf{Mask R-CNN} & \textbf{DeepLab}\\\hline
\textbf{Conv Layers} & 5  & 8 & 57& 132 & 108  \\
\Xhline{1.5pt}
\end{tabular}
}
\caption{CNN models used in our evaluation.}
\label{tbl:networks_res} 
\vspace*{-0.4cm}
\end{table}
}

\subsection{Simulation Methodology}

For performance simulation of \proj{}, we modify GPGPU-Sim 4.0~\cite{bakhoda2009analyzing} and add the systolic mode in the baseline SIMD architecture.
We use GPUWattch~\cite{Leng:2013:GEE:2485922.2485964} and CACTI~\cite{thoziyoor2008cacti} for energy estimation.
We use regular and hybrid models in \Tbl{tbl:networks_res}.

For the GPU-based GEMM implementation, we use NVIDIA's open-sourced and highly optimized CUTLASS library~\cite{cutlass2019}.
In specific, we use the tiling size of $128 \times 128$ and modify it to use the systolic mode as detailed in \Sec{sec:gemm_impl}. 
The convolution layer in CNN models is converted to GEMM through the \benchmark{img2col}. 

\paragraph{Area Overhead}
SMA has little area overhead over the baseline GPU architecture due to the reuse of existing structures. The only significant extra logic is the systolic controller, which has 256B (bytes) storage ($8\times 8$B for $A_{in}$ and $24 \times 8$B for $C_{out}$) and little extra logic. Modern GPU has 256KB register file, 128KB shared memory, and various computations per SM. Therefore, we estimate the overhead is less than 0.1\%.

\begin{figure}[t]
  \vspace{-4mm}
  \includegraphics[width=1\linewidth]{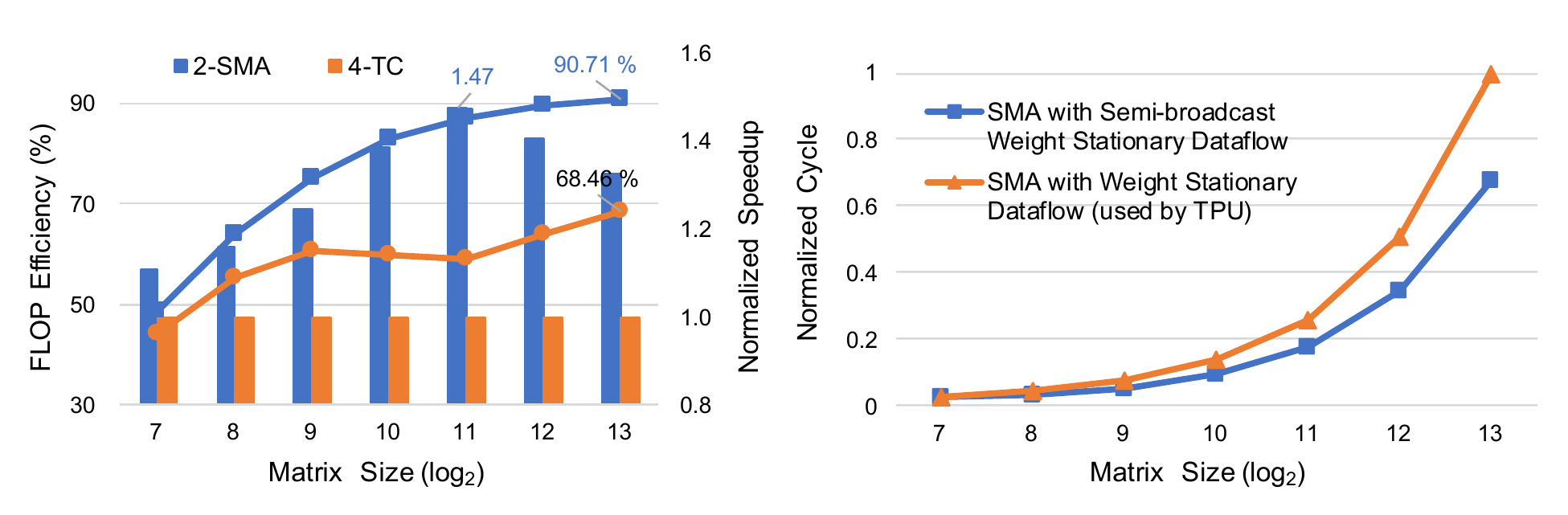}
\vspace*{-0.6cm}
  \caption{\small Iso-FLOP comparison: SMA vs TC (left) and TPU (right).}
  \label{fig:tpu_sma_tc}
\vspace*{-0.4cm}
\end{figure}

\subsection{DNN models}

We evaluate the SMA's efficiency advantages in terms of performance and area.
In specific, we perform an \emph{iso-FLOP} comparison on various data-flows including the SMA, TensorCore and TPU.
We also perform an \emph{iso-area} comparison to demonstrate the advantage of our temporal integration.

\paragraph{Iso-FLOP Comparison}
We first perform the iso-FLOP comparison for the SMA with broadcast weight stationary dataflow,  TensorCore with the dot-product dataflow, and TPU with the weight stationary dataflow. 
Specifically, \Fig{fig:tpu_sma_tc} left plane shows the square GEMM performance in the case of two SMA units (\benchmark{2-SMA}) and four TensorCores (\benchmark{4-TC}) per SM, which both have the same 256 FP16 units.
The \benchmark{2-SMA} achieves 30\% better performance improvement than \benchmark{4-TC} and over 90\% FLOP efficiency (i.e., the ratio of theoretical peak performance) because it eliminates the RF bandwidth limitations.
\Fig{fig:tpu_sma_tc} right plane shows that the TPU dataflow is 20\% - 40\% slower than SMA dataflow because the former has a large amount of shared memory bank conflicts.

\begin{figure}[t]
      \vspace{-3mm}
      \includegraphics[height=0.6\linewidth]{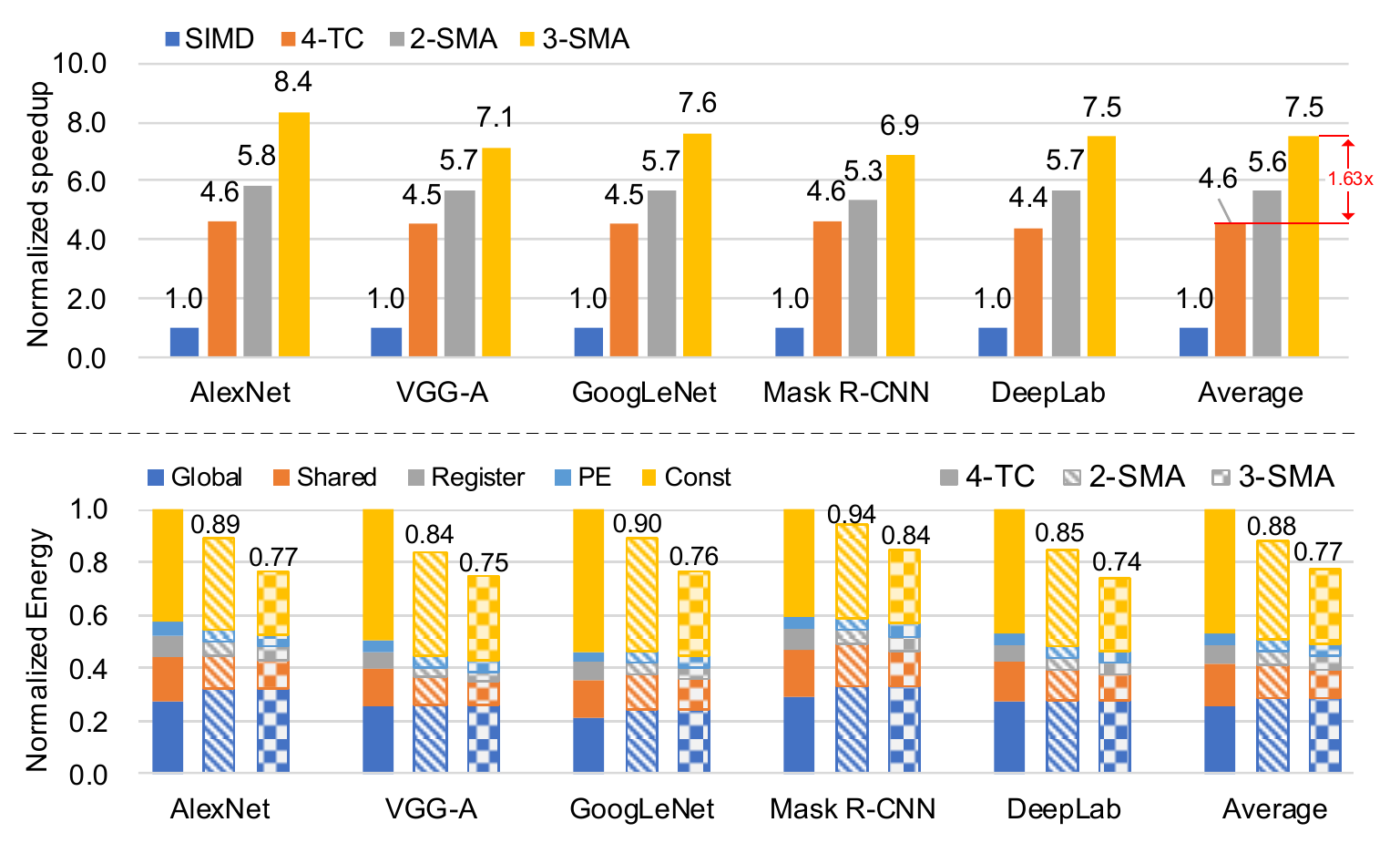}
  \vspace{-5mm}
  \caption{Iso-area comparison for regular and hybrid models.}  \label{fig:networks_res}
  \vspace{-6mm}
    \end{figure}

\paragraph{Iso-Area Comparison}
In the baseline architecture, SIMD units and TC are spatially integrated and the DNN models can only leverage one resource for acceleration.
In contrast, SMA is based on the temporal integration and can use all computation resources.
For the iso-area comparison, we estimate three SMA units (\benchmark{3-SMA}) have the same area with one SIMD unit and two TC, which add up to the area of 384 FP16 units.
\Fig{fig:networks_res} top planes compare performance in various cased on the regular and hybrid models.
The \benchmark{2-SMA} performance is 22\% faster than \benchmark{4-TC} owing to the more efficient dataflow.
The temporal integration leads to 63\% faster \benchmark{3-SMA}.
We also compare the energy consumption between SMA and TC.
As \Fig{fig:networks_res} (bottom) shows, \benchmark{3-SMA} (\benchmark{2-SMA}) consumes 23\% (12\%) less energy than \benchmark{4-TC} on average, where the energy reduction mainly comes from the on-chip memory structures such as register file and shared memory.


    In summary, \proj{} outperforms the baseline GPU in both the performance and energy efficiency for three reasons.
    First, it reduces memory consumption by reusing input/result inside arrays. Under the same memory/register bandwidth, it performs better than the TC, also consumes less memory energy (including shared memory, cache, and RF) and dedicates more energy for the useful computation.
    Second, \proj{} adopts a complex control instruction which mitigates the overhead of instruction fetch/decode.
    Third, the systolic array requires less thread/warp-level parallelism because of the co-ordinated double buffering, which reduces the cache contention.
    The vanilla GPU requires many more threads/thread blocks per SM to hide the memory access latency.

\subsection{End-to-end DL Applications} 

We also evaluate \proj{}'s ability of dynamic resource allocation in the autonomous driving scenario which includes a mixed CNN and non-CNN algorithms.
Prior study shows that it has three major algorithms: detection (DET), tracking (TRA), localization(LOC)~\cite{lin2018architectural}.
The tracking runs after the detection and they are both CNN-based.
The localization runs independently and is not CNN based.
We choose representative DeepLab~\cite{chen2018deeplab}, GOTURN~\cite{held2016learning}, and ORB-SLAM~\cite{DBLP:journals/corr/Mur-ArtalMT15} for them.

\begin{figure}[t]
\vspace*{-0.4cm}
  \minipage{0.5\textwidth}

    \minipage{0.5\textwidth}
    \hspace{0.3cm}
      \includegraphics[height=0.6\linewidth]{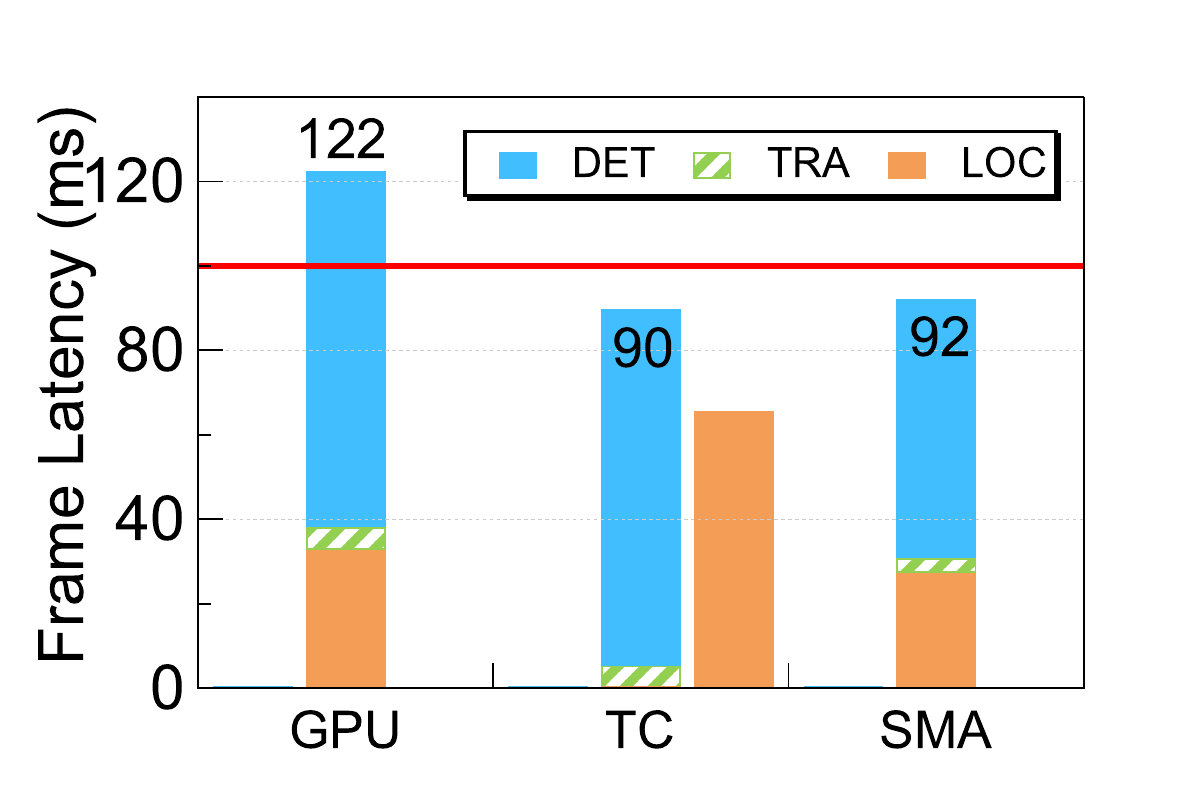}
    \endminipage  
    \minipage{0.5\textwidth}
      \includegraphics[height=0.6\linewidth]{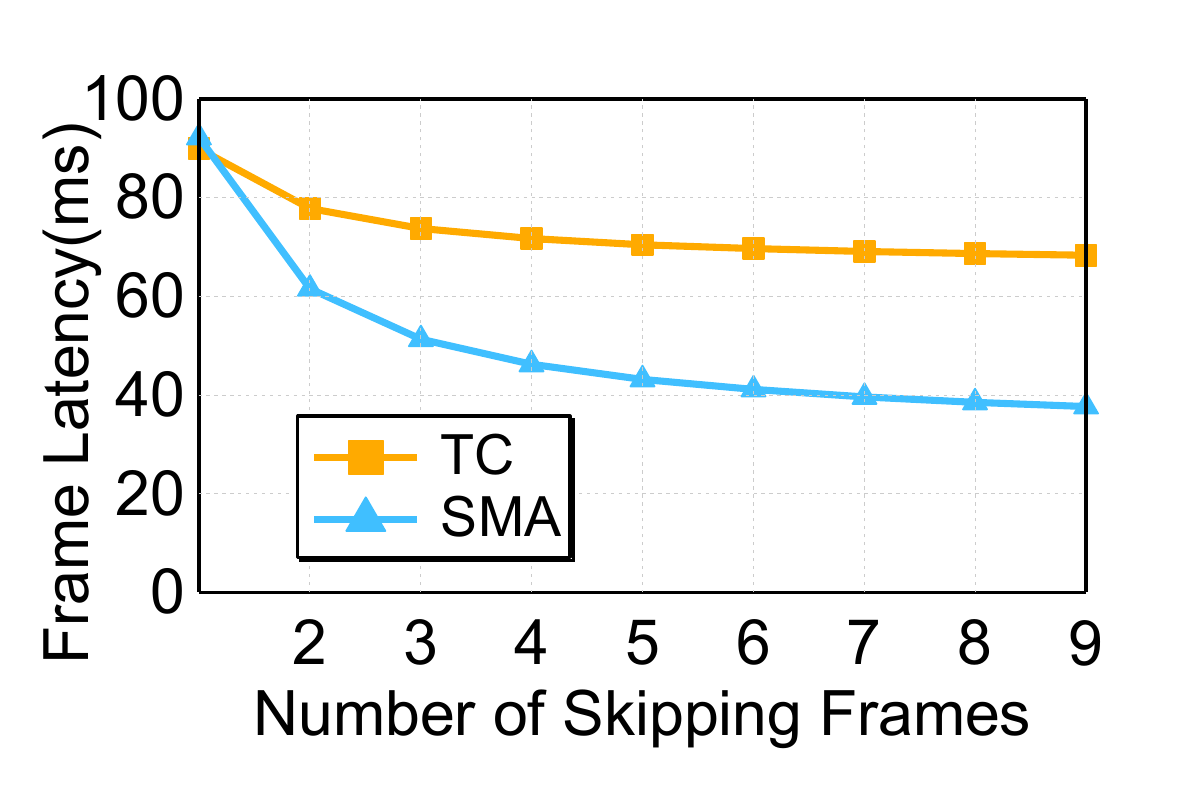}
    \endminipage\hfill
    \vspace*{-0.2cm}
    \caption{End-to-end autonomous driving application results.} 
    \vspace*{-0.5cm}
    \label{fig:endtoend}
    \endminipage
    \end{figure}

The \Fig{fig:endtoend} (left) shows their results in different platforms.
Three algorithms can occupy the entire GPU so the frame latency equals the sum of each algorithm.
The GPU exceeds the 100~ms single frame latency target owing to the slow CNN performance.
The execution on \proj{} is similar but meets the latency target because of the faster CNN performance.
The TC has a similar latency of \proj{}, but with DET and TRA running sequentially on the TC, and LOC running on the GPU in parallel.
However, these results are based on running object detection and tracking on every frame.
Prior work has suggested only running the detection every $N$ (e.g. 4) frames and relying on the tracking for every frame does not impact the final accuracy~\cite{DBLP:journals/corr/abs-1803-11232}.
This dynamic optimization creates uneven demand for CNN computation which \proj{} can accommodate to reduce the frame latency.
\Fig{fig:endtoend} (right) shows with $N = 4$, \proj{} can reduce the frame latency by almost 50\%.

\section{Related Work}
\label{sec:related}

There are many recent works on designing for deep learning accelerators~\cite{chen2016diannao, jouppi2017datacenter, 7551407}.
Prior work also tried a programmable acceleration through FPGA and CGRA~\cite{Voitsechov:2014:SMF:2665671.2665703,7446051}.
Those work, together with general-purpose architecture and ASICs, represent different points in the trade-off curve between generality and efficiency.
Our work integrates the two extreme points in the same architecture.  
Recent Volta GPU spatially integrates the TensorCore accelerator~\cite{raihan2019modeling} while we temporally integrate the SIMD architecture with the systolic array.



\section{Conclusion}

We develop a simultaneous multi-mode architecture (\proj{}) with lightweight integration approach on GPU to achieve high programmability and energy-efficiency. 
Using the systolic array, \proj{} significantly improves the performance and reduce the energy.
It also has the same characteristics as GPU and maintains the GPU's programmability, configurability, and generality for fast-evolving DNN workloads.

\vspace{0.35cm}
\noindent \textbf{Acknowledgement}
\vspace{0.05cm}
We thank the anonymous reviewers for their constructive feedback for improving the work.
This work was supported by National Key R\&D Program of China (2019YFF0302600), the National Natural Science Foundation of China (NSFC) grant (61702328, 61832006, 61729202, and U1636210), CCF-Tencent Open Fund. Any opinions, findings, and conclusions in this paper are those of the authors only and do not necessarily reflect the views of our sponsors. 

{\tiny
\bibliographystyle{abbrv}
\bibliography{references}
}

\end{document}